\documentclass[%
 reprint,
superscriptaddress,
frontmatterverbose, 
preprintnumbers,
 amsmath,amssymb,
pra,
floatfix,
]{revtex4-1}

\usepackage{graphicx}
\usepackage{dcolumn}
\usepackage{bm}
\usepackage{physics}
\usepackage{amsmath}
\usepackage{amsthm}
\usepackage{hyperref}

\begin{document}


\title{Stochastically Realized Observables for Excitonic Molecular Aggregates}

\author{Nadine C Bradbury}
\email{nadinebradbury@ucla.edu}
\affiliation{Department of Chemistry and Biochemistry, University of California, Los Angeles, California, 90095, USA}
\author{Chern Chuang}
\affiliation{ Department of Chemistry, University of Toronto, Toronto, Ontario CA}
\author{Arundhati P Deshmukh}
\affiliation{Department of Chemistry and Biochemistry, University of California, Los Angeles, California, 90095, USA}
\author{Eran Rabani}
\affiliation{Department of Chemistry, University of California and Material Science Division, Lawrence Berkeley National Laboratory, Berkeley, California 94720, USA}%
\author{Roi Baer}
\affiliation{Fritz Haber Center for Molecular Dynamics, Institute of Chemistry, The Hebrew University of Jerusalem, Jerusalem 91904, Israel}
\author{Justin R Caram}
\email{jcaram@chem.ucla.edu}
\affiliation{Department of Chemistry and Biochemistry, University of California, Los Angeles, California, 90095, USA}
\author{Daniel Neuhauser} 
\affiliation{Department of Chemistry and Biochemistry, University of California, Los Angeles, California, 90095, USA}

\date{\today}

\begin{abstract}
    We show that a stochastic approach enables calculations of the optical properties of large 2-dimensional and nanotubular excitonic molecular aggregates. Previous studies of such systems relied on numerically diagonalizing the dense and disordered Frenkel Hamiltonian, which scales approximately as $\mathcal{O}(N^3)$ for $N$ dye molecules. Our approach scales much more efficiently as  $\mathcal{O}(N\log(N))$, enabling quick study of systems with a million of coupled molecules on the micron size scale. We calculate several important experimental observable including the optical absorption spectrum and  density of states, and develop a stochastic formalism for the participation ratio. Quantitative agreement with traditional matrix diagonalization methods is demonstrated for both small- and intermediate-size systems. The stochastic methodology enables the study of the  effects of spatial-correlation in site energies on the optical signatures of large 2D aggregates. Our results demonstrate that stochastic methods present a path forward for screening structural parameters and validating experiments and theoretical predictions in large excitonic aggregates.
\end{abstract}

\maketitle


\section{Introduction}
    Excitonic molecular aggregates are ubiquitous in molecular electronics and photosynthetic light harvesting systems.\cite{Brixner2017} In these systems, coupling among transition dipole moments enables collective interactions with the electromagnetic field.  Long-range dipole-dipole interactions induce complex and tunable photophysical properties, such as superradiance,\cite{Doria2018,Spano1989} exchange narrowing,\cite{Malyshev1999} strong polarization dependent behavior,\cite{Spitz2002} and long-range transport properties.\cite{Fidder1993, Caram2016,Pandya2019} Particular applications of these materials are as photo-emitters and antennas, and they are highly desired for numerous technological, medical, and biological imaging applications.\cite{Chen2019,Hansen2008,Bouit2007,Wei2020} Given the interest in the optical properties of these dye aggregates, approaches to rationalize and control excitonic properties aggregation are a subject of recent research.\cite{Bricks2015,Hestand2017,Deshmukh2019} Thoroughly testing design principles new aggregate complexes is difficult, as the traditional Frenkel exciton matrix diagonalization approach becomes prohibitively expensive for large systems.
   
    Experimental and theoretical exploration of the optical properties of molecular aggregates is nearly a century old.\cite{Jelley1936,Davydov1964, Kasha1963} In recent years, advances in chromophore design and self assembly has allowed for the creation of tubular and 2D aggregates which have potential as excitonic antennae.\cite{Bondarenko2020, Deshmukh2019,Chuang2019} However, the slow convergence of the $r^{-3}$ dipolar coupling necessitates calculating  band structures for extremely large systems.\cite{Chuang2016} This is exacerbated in 2-D and quasi-2D tubular systems for which the number of sites grows non-linearly with system size.  Without methods which treat large systems, computational studies are limited to diagonalizing Hamiltonians representing a few thousand dye monomers, and  observed localization effects of disorder depend on the size of the calculation.\cite{Bondarenko2020,Didraga2004}  Larger systems are approached analytically with highly limiting assumptions, such as nearest-neighbor interactions or zero disorder.  Probing 2-D aggregates at the length scales observed experimentally (microns),\cite{Eisele2012} stochastic methods provide an appealing alternative to insurmountable diagonalization tasks.
    
    The idea of calculating the density of states through stochastic expectation values of a polynomial approximation for the delta density operator is well established.  Its foundations go back to Lanczos in 1950,\cite{lanczos1988applied} but the essential algorithm has been significantly refined in the 1970s and 1990s in the fields of nuclear physics and quantum chemistry.\cite{Gautschi1968,Gautschi1970,Sack1971,Wheeler1972,Blumstein1973,Drabold1993,SILVER1994-PRB,Wang1994-PRB}  Based off its numerical accuracy and ease of implementation, it has become a staple method for computation of large quantum systems, and is now often known as the kernel polynomial method.\cite{Weisse2006} 
    To date, similar stochastic methods have been applied to complex excitonic systems with similar computational requirements as molecular aggregates, like quantum dots.\cite{Wang1994-PRB, Baer2012-NL}
    
    The stochastic approach for calculating the density of states is highly suitable for our specific case of dipole-coupled dyes in ordered 2D planar or tubular systems.  This is because the effective exciton Hamiltonian that needs to be diagonalized has a special  form, i.e., the coupling between sites depends only on the distance between them.  This makes it very efficient to calculate, in a quasi-linear scaling, the required kernel moments using convolution.  An additional advantage is that the method is automatically suitable for including many kinds of energy disorder, without additional cost, as the averaging over the different disorder is included as part of the stochastic averaging of the moments.
    
    Following earlier work on the stochastic resolution of the identity (SIR),\cite{Baer2012,Neuhauser2012,Neuhauser2013,Baer2013} we show that, in addition to the calculation of the density of states,  the stochastic approach enables the calculation of a further quantity that measures  exciton delocalization.  This quantity, the  participation ratio,\cite{Thouless1974}  is obtained here with the same overall scaling as the density of states.
    
    The overall approach presented here enables extremely fast screening of aggregate geometries and disorder, unlocking rapid computation of experimentally relevant parameters optical parameters.

\section{Computational Methods}
\subsection{Hamiltonian, spectra and participation ratio}
    We study here the  Frenkel-Exciton Hamiltonian for interacting molecular chromophores,\cite{Davydov1964}
    \begin{equation}
    H = \sum_n \epsilon_n\ket{n}\bra{n} + \sum_{nm} J(\bm{n}-\bm{m})\ket{n}\bra{m}, \label{hamiltonian}
    \end{equation}
    where $n$ represents the site basis of an exciton localized on a single monomer. $\epsilon_n$ are the on site excitation energies. We set the average monomer excitation energy to 0 artificially to study specifically the effects of aggregation.
    
    The primary tool by which optical properties of excitonic molecular aggregates are usually studied is through explicit construction and diagonalization of the Frenkel Hamiltonian matrix. A variety of different off-diagonal coupling functions may be used to capture the transition dipole coupling or charge transfer effects .\cite{Hestand2017,volkhardmay2011,Merrifield1961, Hestand2016} The important optical properties are then assessed through several quantities defined below: optical absorption, density of states, and participation ratios. 
    
    The optical absorption coefficient (abbreviated here as optical absorption) is 
    \begin{align}
        A(\omega) &= \sum_i
        \left(\bm{E}\cdot \bm{\mu}\right)^2  \delta(\omega-\epsilon_i) \label{absorption}\\
        &= \sum_i \left| \bra{\psi}\ket{\phi_i}\right|^2\delta(\omega-\epsilon_i).
    \end{align}
Here, $\varepsilon_i$ and $\ket{\phi_i}$ are the eigenvalues and eigenvectors of $H$. $\bm{\mu}$ is the dipole moment operator, and $\bm{E}$ is the electric field polarization. For a system small relative to the wavelength of the absorbed radiation, the so called optically bright state $\ket{\psi}$ would be the $\bm{k} = 0$ state, with elements 
    \begin{equation}
        \braket{n}{\psi} = \bm{\mu_n}\cdot \bm{E} \label{bright_eq}
    \end{equation}
    where  $\bm{\mu_n}$ is now refers to the dipole vector of an individual monomer. The $\bm{k} = 0$ state is the most studied, so it is what we restrict to in this paper, though the systems are large enough that full consideration beyond the dipole limit may be appropriate for future work. The stochastic method can easily be extended to do the full absorption through the addition of a spatial filter (See Appendix \ref{fullabsorb}).
    
    The density of states is, 
    \begin{equation}
        \rho(\omega) = {\rm{Tr}}[\delta(H- \omega)] = \sum_i \delta(\varepsilon_i -\omega),
    \end{equation}
    and the participation ratio is defined as,
    \begin{equation}
        \mathcal{P}(\omega) = \frac{\rho(\omega)}{K(\omega)}, \label{ipr-eq}
    \end{equation}
    where
    \begin{equation}
        K(\omega)\equiv\sum_i \delta(\varepsilon_i - \omega)\sum_n |\braket{n}{\phi_i}|^4.
        \label{Kw}
    \end{equation}
Average aggregate properties should be estimated by many realizations of the Hamiltonian with different  disorder. This additional cost further reduces the maximum practical aggregate size that can be studied using direct diagonalization.
    
\subsection{The Chebyshev expansion}
As mentioned, in this paper we use a stochastic trace of the delta density operator to retrieve the density of states. Before we can take the trace, the delta function is first numerically implemented with Gaussian regularization.\cite{Weisse2006} The regularized density operator is defined through the Chebyshev polynomial expansion \cite{Kosloff1988}
    \begin{equation}
        F(\omega) = 
            \frac{1}{\gamma\sqrt{\pi}}e^{-(H-\omega)^2/\gamma^2} = \sum_{\ell=0}^{N_{Chebyshev}} c_\ell(\omega) T_\ell(H') \label{cheby}
    \end{equation}
and of course in the small $\gamma$ limit, $F(\omega)\to \delta(H-\omega)$.  Here, $T_\ell(H')$ is the $\ell$'th Chebyshev polynomial of a linearly scaled Hamiltonian $H'=(H-{\bar{h}})/\Delta H $  constructed so that its  eigenvalues are within the interval $[-1,1]$;  $ \bar{h}$  is an estimate for the center of the spectrum of $H$, and $2\Delta H$ is an upper bound for its spectral width. $N_{Chebyshev}$ is the required number of  Chebyshev polynomials, which is proportional to $\Delta H / \gamma$.
    
As discussed later, the coupling in the Hamiltonian only depends on the difference of position between sites, so if there is no disorder $\Delta H$ can easily be shown to be given from a 2D Fourier transform of the elements in the Hamiltonian.  Accounting for the effect of the disorder, we enlarge the spectral width by a factor to ensure the stability of the Chebyshev expansion. 
    
The scalar Chebyshev coefficients are calculated using the transform $\theta = \cos^{-1}(x)$. 
    \begin{align}
        c_\ell(\omega) &=
         \frac{1}{\sqrt{\pi}\gamma}
        \int_{-\infty}^\infty dx \frac{e^{-(\Delta H x-\bar{h} -\omega)^2/\gamma^2}T_\ell(x)}{\sqrt{1-|x|^2}} \\
        &=   \frac{2-\delta_{\ell,0}}{\sqrt{\pi}\gamma} \int_{0}^{2\pi} d\theta e^{-(\Delta H  \cos\theta-\bar{h} -\omega)^2/\gamma^2} e^{i\ell\theta} \label{coefs}.
    \end{align}
    The coefficients are then calculated via Eq. (\ref{coefs}) using a fast Fourier Transform (FFT). 
    
\subsection{Absorption Spectrum}
    From Eq. (\ref{absorption}), the absorption spectra is calculated with the Chebyshev expansion using only the optically absorbing bright state
    \begin{equation}
        A(\omega) = \bra{\psi}F(\omega)\ket{\psi}.
    \end{equation}
    This expectation value can be calculated for each coordinate of the electric field, $\bm{E}$, and therefore a bright state along each coordinate can be defined via (Eq. (\ref{bright_eq})). This gives the dichroism response.
    
\subsection{Stochastic Density of States}
    
To take the trace of the moments operator, a stochastic state is introduced, which Monte-Carlo samples a complete basis for $H$ (see Ref.\cite{Wang1994-PRB}). The stochastic excitation has a random $\pm 1$ amplitude at each site,  $\zeta(n) \equiv \braket{n}{\zeta} = \pm 1$. Thus, the DOS is calculated directly as 
    \begin{equation}
        \rho(\omega) = \Bigl\{\bra{\zeta}F(\omega)\ket{\zeta}\Bigr\}=
        \sum_\ell c_{\ell}(\omega) R_\ell,
        \label{rhokernels}
    \end{equation}
where curly brackets are introduced to represent a classical expectation value over the random excitations, and the kernels are 
    \begin{equation}
        R_\ell \equiv \Bigl\{ \braket{\zeta}{\zeta^\ell}\Bigr\},
        \label{kernels}
    \end{equation}
where we defined the Chebyshev vectors, 
\begin{equation}
  \ket{\zeta^{\ell}}\equiv T_\ell(H')\ket{\zeta}
\end{equation}
obtained recursively by the usual Chebyshev recursion relation, $\ket{\zeta^{\ell}}=
2H'\ket{\zeta^{\ell-1}} - 
\ket{\zeta^{\ell-2}}.$

The proof of Eq. (\ref{rhokernels}) follows once we expand the random vector in terms of the site basis set
    $ \ket{\zeta}=\sum_i \zeta(n) \ket{n}$,
    and use 
    $\{ \zeta(n) \zeta(m) \} = \delta_{nm}$.
    This approach to the density of states converges rapidly with the line broadening parameter $\gamma$, and is memory efficient, as one stores only the kernels  and coefficients.
    
\begin{figure}
    \centering
    \includegraphics[width = 0.45\textwidth]{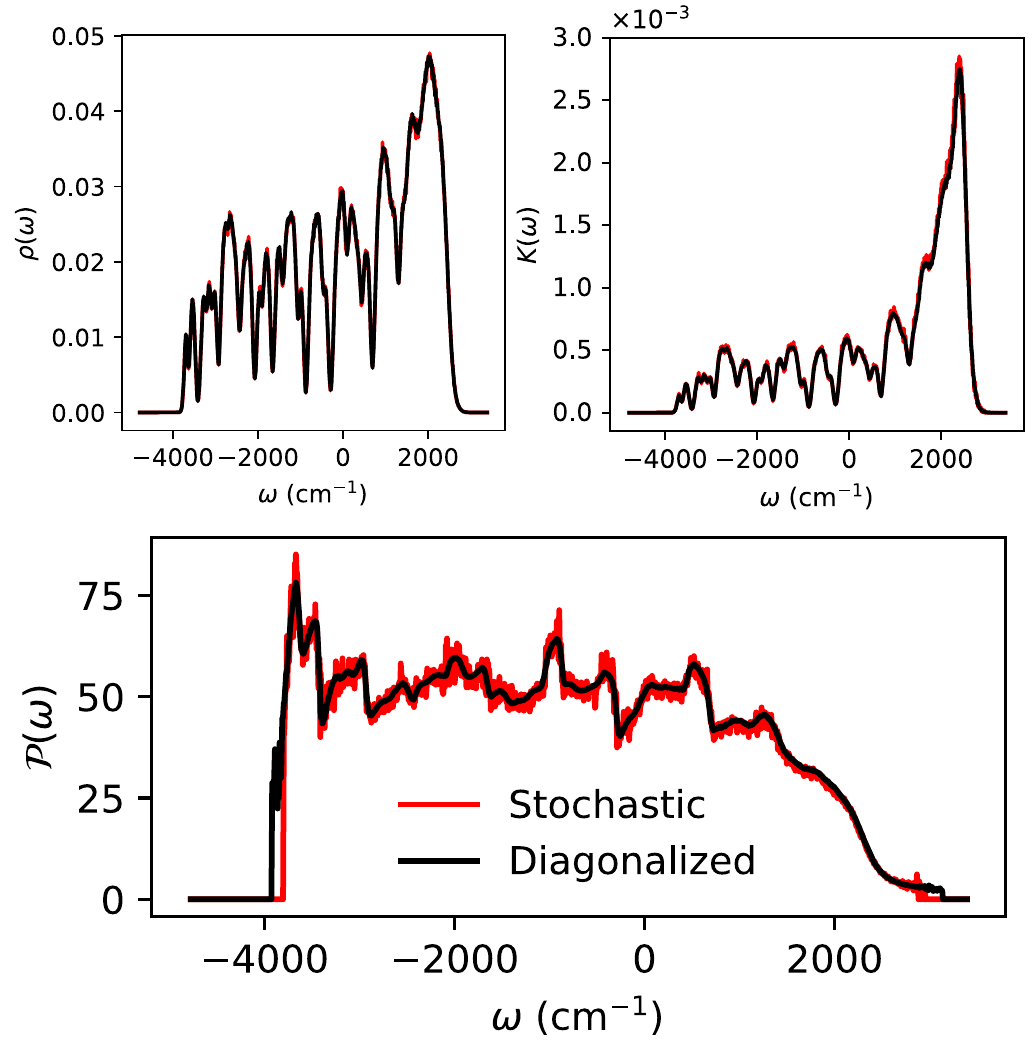}
    \caption{Demonstration of the accuracy of the stochastic resolution of the participation ratio. Top left is the density of states, top right the denominator of the participation ratio $K(\omega)$, and the participation ratio is shown at the bottom. A small system of $N=15\times 9=135$ monomers is simulated here with $N_{stochastic} = 5\times 10^5$ samplings (of $\zeta$ and the noisy diagonal energies, with disorder $\sigma = 400 \, \mathrm{cm^{-1}}$ and no site-to-site correlation of the diagonal energies). In accordance with the small $\gamma$ limit necessary for the accuracy of the ratio, we used $\gamma = 2 \, \mathrm{cm^{-1}}$ and $N_{Chebyshev} = 16384$. The very high-wavelength fluctuations are due to stochastic error, and can be flattened either by more samplings or by explicit smoothing.}
    \label{ipr.small}
\end{figure}

\subsection{Stochastic Participation Ratio}
To have a fully stochastic expression for the participation ratio,  we need a stochastic formalism that samples the fourth power of the eigenvectors accurately, i.e., the denominator of Eq. (\ref{ipr-eq}). This is done here analogously to the stochastic estimation of the exchange and MP2 energies.\cite{Baer2012, Neuhauser2012,Ge2013,Neuhauser2015}

For a given broadening parameter, $\gamma$, we first pick two independent random vectors, $\ket{\zeta}$ and $\ket{\xi}$, each defined similarly to the random vector in the previous section with $\pm 1$ at each grid site. We then define filtered-vectors:
\begin{equation}
\ket{{\bar\zeta}(\omega)}\equiv F^{1/4}(\omega)\ket{\zeta}, \,\,\,\,\,
\ket{\bar\xi(\omega)}\equiv F^{1/4}(\omega)\ket{\xi},
\end{equation}
where
$
F^{1/4}(\omega)= \frac{1}{\gamma^{1/4}\pi^{1/8}} e^{-(H-\omega)^2/4\gamma^2}
$. These vectors are calculated using Eq. 
(\ref{cheby}), i.e.,
\begin{equation}
\ket{{\bar\zeta}(\omega)}=
\sum_{\ell} \bar{c}_\ell(\omega) \ket{\zeta^{\ell}}.
\end{equation}
Here, $\bar{c}_\ell(\omega) $ are the Chebyshev coefficients associated with $F^{1/4}(\omega)$. Given the filtered vectors, the stochastic expression for the denominator in Eq. (\ref{ipr-eq}) is
$  K(\omega)=\lim_{\gamma\to 0} K_\gamma (\omega)$ where
\begin{equation}
    K_\gamma(\omega)=
    \Bigl\{ 
    \sum_n 
     \bigl|\braket{n}{\bar\zeta(\omega)} 
     \braket{n}{\bar\xi(\omega)}\bigl|^2 
       \Bigr\}.
       \label{Kgw}
\end{equation}
To prove this expression, we first formally expand  each vector in terms of the complete basis of eigenstates of $H$, 
\begin{equation}
    \ket{\zeta} = \sum_i a_i \ket{\phi_i}, \,\, \ket{\xi} = \sum_j b_j \ket{\phi_j},
\end{equation}
where $a_i\equiv \braket{\phi_i}{\zeta}$, etc. While the coeficients $a_i$ do not have a closed form like the elements of $\ket{\zeta}$, they remain uncorrelated ($\{a_i a_j\} = \delta_{ij}$) due to their construction from $\ket{\zeta}$. We also define
$$
f_i(n) = \bra{n}F^{1/4}(\omega)\ket{\phi_i}
=\delta^{1/4}(\epsilon_i-\omega) \phi_i(n)
$$
without explicitly denoting the $\omega$ dependence of $f_i(n)$.  

Plugging to the expression for $K_{\gamma}(\omega)$, we get
\begin{equation}
    K_\gamma(\omega)  
    = \sum_n \sum_{ijkl} 
    \bigl\{ a_i a_j b_k b_l \bigr\} f_i(n) f_j(n) f_k(n)f_l(n) 
\end{equation}
and using 
\begin{equation}
    \{ a_i a_j b_k b_l \} = \{a_i a_j\}\cdot\{b_k b_l\} = \delta_{ij}\delta_{kl},
\end{equation} 
leads to
\begin{equation}
\begin{split}
 &K_\gamma (\omega)=\sum_n \left( \sum_i\left( f^{1/4}_i(n)\right)^2\right)^2 \\
    &=
    \frac{1}{\gamma \sqrt{\pi}}
    \sum_n \sum_{ij}  e^{-(\varepsilon_i-\omega)^2/2\gamma^2}
    e^{-(\varepsilon_j-\omega)^2/2\gamma^2} 
    \braket{n}{\phi_i}^2 \braket{n}{\phi_j}^2
    \end{split}
\end{equation}
and taking the limit  $\gamma\to 0$ and in the limit of any disorder to break eigenstate degeneracies,
\begin{equation}
\begin{split}
     K(\omega) &=  \lim_{\gamma \to 0 }
      \frac{1}{\gamma \sqrt{\pi}}
      e^{-(\varepsilon_i-\omega)^2/2\gamma^2}   e^{-(\varepsilon_j-\omega)^2/2\gamma^2}  \braket{n}{\phi_i}^2 \braket{n}{\phi_j}^2  \\ &=   \delta_{ij} \delta(\varepsilon_i-\omega) . \braket{n}{\phi_i}^4,
    \label{limexpexp}
    \end{split}
\end{equation}
finally leading to Eq. (\ref{Kw}), as stipulated.

The  estimate for the denominator in the participation ratio, Eq. (\ref{Kgw}), converges well statistically, since it is an average of positive definite quantities, but its $\gamma$ dependence relates to the system size and disorder strength:
\begin{itemize}
    \item For small $N$ the accuracy of the overall participation ratio depends  much more strongly on reaching the small gamma limit than for the density of states alone, as shown in Fig. \ref{ipr.small}.
    \item In contrast, for large $N$ (beyond $10^4$) the participation ratio converges rapidly with the number of stochastic samples and with gamma, due to self-averaging and the fact that different states have little spatial overlap.  Put differently, the $i\ne j$ terms in Eq. (\ref{limexpexp}) become minuscule due to the reduced overlap of eigenvectors for large systems, not just due to being a sum over spatially destinct Gaussians at small $\gamma$. For further details, see Appendix \ref{interpolated}. 
\end{itemize}
 
A complication in the participation ratio calculation is  that  memory-constraints rather than CPU time usually limit the fesible system size, $N$. This is due to the need to store the set of $\ket{\bar{\zeta}(\omega)}$ vectors, of size $N_\omega\cdot N$, which for a large system quickly reaches gigabytes of CPU memory per core if significant resolution across the band is desired.

\subsection{Choice of Coupling Function \label{convolution}}
\begin{figure*}
    \centering
    \includegraphics[width=0.8\textwidth]{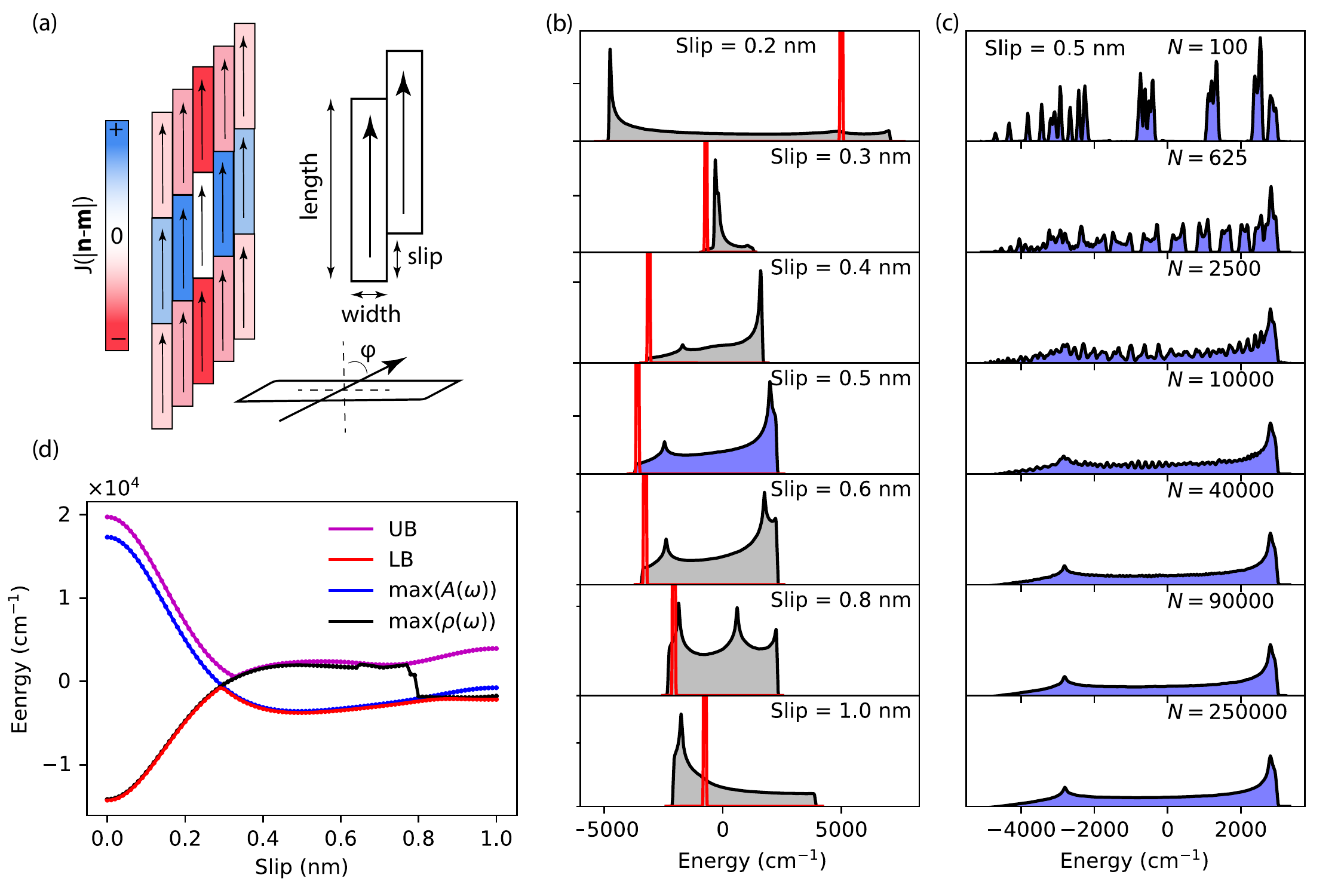}
    \caption{(a) Diagram of section of 2D planar aggregate. The relative coupling strengths for near neighbors of a given site are shown by different colors.  (b) DOS (grey) and Absorption spectra (red) for various slip values. Standard geometry parameters of length and width of 2nm and 0.4nm respectively are used for all aggregates (see Appendix \ref{geometry}).\cite{Chuang2019} (c) Examples of the Slip=0.5 planar DOS for different system sizes. As with all calculations, we have done perfect toriodal lattice boundary conditions. Fluctuations in the center of the DOS still appear at system sizes of about 10,000. Further driving the need to simulate big systems, or use artificial boundary conditions. Mild disorder of 50 cm$^{1
    }$ is additionally used to help smooth out the DOS. (d) Scan across 100 slip values, showing the upper (UB) and lower (LB) band edges as well as the position of the absorption peak and position of the tallest Van Hove peak. 
    \label{slipfig}}
\end{figure*}

An underlying key element of the iterative stochastic approach is the use of a Hamiltonian with off diagonal components that depend only on the distance between sites, or difference of indices, and the use of a perfect lattice. This makes it feasible to apply the Hamiltonian on a vector with quasi-linear cost.  Specifically, here we use the point dipole approximation,
\begin{equation}
    J(\bm{n}-\bm{m}) = \frac{\bm{\mu_n}\cdot\bm{\mu_m}}{|\bm{r_{nm}}|^3} -3 \frac{(\bm{\mu_n}\cdot\bm{r_{nm}})(\bm{\mu_m}\cdot\bm{r_{nm}})}{|\bm{r_{nm}}|^5} \label{coupling-eq}
\end{equation}
with $\bm{r_{nm}} = \bm{r_n} -\bm{r_m}$. Eq. (\ref{coupling-eq}) is applied to aggregates with both planar and tubular geometry.\cite{Chuang2019,Didraga2002, Didraga2004} Fig. \ref{slipfig} contains a diagram showing how the coupling is constructed from the aggregate geometry. System geometry is further discussed in Appendix \ref{geometry}.

For perfect toroidal boundary conditions, the Frenkel Exciton Hamiltonian, Eq. (\ref{hamiltonian}), forms a block circulant matrix, with block sizes $N_x$ and $N_y$, and is thus diagonalized by a 2D Fourier Transform.\footnote{This is formally true if $N_x$ and $N_y$ are odd, due to the even nature of the coupling functions. For sufficiently large $N_x$ and $N_y$ the phase introduced by an even number of samples is suppressed below machine error. Thus for small systems, products of small odd primes are suggested, but divisors of 2 are acceptable for macroscopic systems.}  At sufficiently large block sizes, perfect periodic boundaries (toroidal) do not impose an issue with self coupling. Multiplication by a block circulant matrix is done by the two dimensional convolution theorem,
\begin{align}
    b_j &= (H a)_j = \sum_i H_{ji} a_i = \epsilon_ja_j + \sum_i J(\bm{i}-\bm{j})a_i \\
    &= \epsilon_ja_j + \mathcal{F}^{-1} [ \tilde{J}(\bm{k}) \mathcal{F}[a]]
\end{align}
where $\mathcal{F}$ represents the Fourier transform. Open boundary conditions, such as in the most recent computational work on tubular aggregates,\cite{Bondarenko2020} can be achieved via zero-padding of the coupling matrix.
\begin{figure}
    \centering
    \includegraphics[width = 0.45\textwidth]{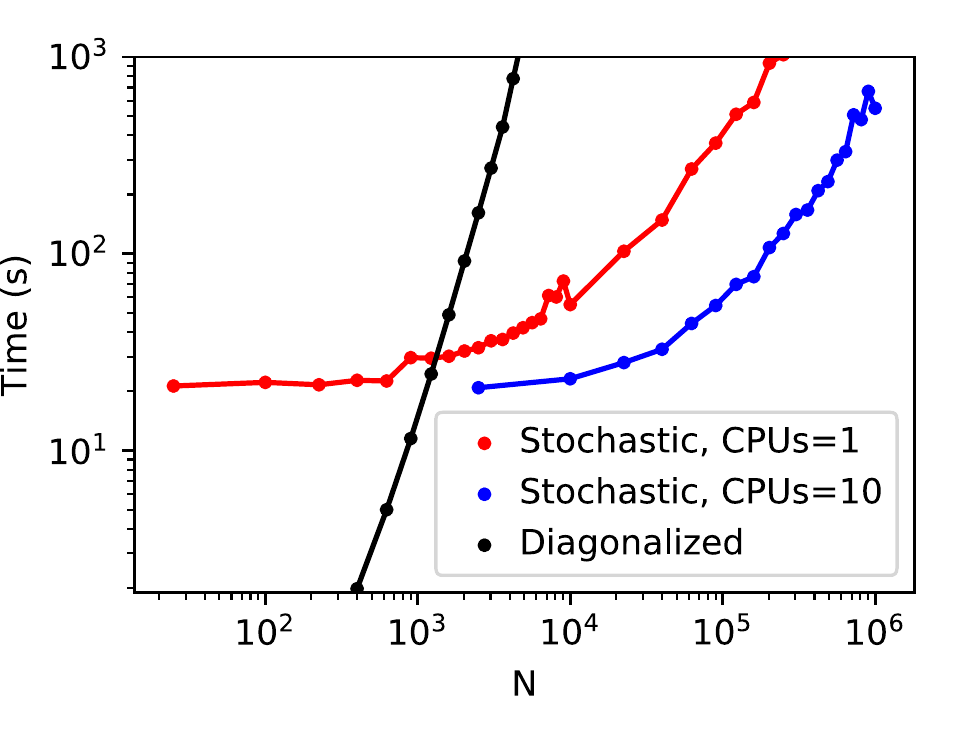}
    \caption{Timing test on the program that calculates absorption and density of states, comparing a single core (red) and ten cores (blue). For very small systems, the time is approximately constant, and then scales like $\propto N\log(N)$ for larger systems. For all calculations $N_{stochastic} = 10$ and $N_{Chebyshev} = 4096$, which is enough to converge the integral density of states to the exact value of $N$. Small wiggles in the timing are due to the different relative efficiency of the FFT package used, FFTW3,\cite{Frigo2005} at different array sizes. For the diagonalization method, the full dimension $N\times N$ hermitian Hamiltonian matrix is constructed, diagonalized, and the density of states is calculated from the eigenvalues. Only a single instant of diagonalization (no disorder) is considered here. All times were recorded with the Linux `time' command on an AMD EPYC 7452 32-Core Processor at 3 GHz. }
    \label{time}
\end{figure}

\subsection{Overall Algorithm Scaling}

The main numerical CPU cost is due to the repeated application of the Hamiltonian ($N_{Chebyshev}$ times) and specifically the convolutions parts, costing in FFT about   $10 N\log_2(N)$ each time.  In addition, when we calculate the participation ratio we need to accumulate frequency-resolved Chebyshev vectors. Thus the total cost is approximately
\begin{equation}
    N_{opperations} = N_{Stochastic}N_{Chebyshev}N\Bigl(10\log_2(N) + N_{\omega}\Bigr) 
\end{equation}The Monte-Carlo sampling is done in parallel on each node (using MPI) with every node starting from a different random excitation.

The scaling is exemplified in Fig. \ref{time}. Both $N_{Chebyshev}$ and $N_{Stochastic}$ do not scale up with $N$, so the algorithm scales quasi-linearly with $ N$.  Specifically:
\begin{itemize}
\item
$N_{\omega}$ is fixed for constant resolution, since $\Delta H$ does not really scale with system size. 
\item
$N_{Chebyshev}$ is about  $5\frac{\Delta H}{\gamma} \sim 2000 - 8000$. 
For most of these aggregate systems without extreme disorder, the spectral width is on the order of about $10^5 \, \mathrm{cm^{-1}}$, while the spectral line width, $\gamma$, need only be about as good as one could achieve experimentally,  i.e., $\approx 1\, \mathrm{cm^{-1}}$ or larger. Note that our choice of using the most studied point dipole coupling function is known to overestimate nearest-neighbor couplings, and thus the spectral width.\cite{Deshmukh2019} One would expect a decrease in the number of coefficients with more sophisticated or system specific coupling functions.
\item
In the regime of disorder studied, $N_{Stochastic}$ does not scale with system size. In fact, due to self averaging in large systems the error goes like $\propto 1/\sqrt{N N_{Stochastic}}$,\cite{Wang1994-PRB,Weisse2006}, so $N_{Stochastic}$ is reduced commensurately with the system size.
\end{itemize}

\subsection{Disorder}
A key feature of a Monte-Carlo based approach is the ability to vary multiple input parameters at once and still sample the general spectrum. As such, disorder poses no new additional cost to the algorithm. We study the most common kind of disorder, diagonal site disorder $\epsilon_i$.  Latter papers will study the effects of disorder in the dipole direction and of deviations from the ideal lattice positions.

The simplest model of diagonal-site disorder is non-correlated noise, usually via a normal distribution of standard deviation $\sigma$. More sophisticated models introduce correlations into the site disorder. Specifically, the study of the effects of exponentially correlated site disorder is known as Knapp's model in molecular aggregates.\cite{Hestand2018} Knapp suggested that correlation in disorder may be important in organic molecular aggregates, modeling lattice defects and mixtures with glasses, and strong low-frequency exciton-phonon coupling where there is no resolvable vibronic structure.\cite{Knapp1984} Such a strongly coupled low energy phonon mode was indeed recently observed in light-harvesting nanotube aggregates, prompting new interest in correlation in two dimensional and tubular aggregates.\cite{Pandya2018}

Computational work on correlated disorder has a rich literature in one-dimensional systems,\cite{Izrailev2012,Knoester1993, Spano2005,Spano2009} and recent work on two-dimensional nearest-neighbor lattices.\cite{deMoura2010} Correlation has yet to be studied in large non-biological aggregate systems, or in two dimensional systems with full coupling. Studies of correlated disorder in 1D and higher dimensions have long suggested that localized states may exist at all levels of disorder.\cite{Dunlap1989,Fidder1991} 

In photosynthetic systems, there are common claims that small-scale correlated fluctuations may effect their emissive properties. The most heavily studied model is the Fenna–Matthews–Olson (FMO) complex, in which long lived quantum coherences between chromophores suggest relevant spatial correlations between chromophores.\cite{Lee2007,Fidler2012,Panitchayangkoon2010} Similarly long lived quantum coherences due to spatial correlation in multi-exciton dynamics have been observed in quantum dots.\cite{Caram2013,Cassette2015,Pal2017} These experiments all suggest relevant correlation length scales of sub-nm scale or smaller.

There have been studies using mixed quantum and classical photosynthetic systems showing the effects of intersite correlation .\cite{Olbrich2011} Few-state quantum mechanical models, similar to the calculations done here (but for much smaller scales), show large influence of even small correlations between chromophores, and agree qualitatively with the experimentally observed lifetimes and coherences.\cite{Abramavicius2011, Huo2012, Rebentrost2009} Without an experimentally solved system structure and the difficulty in treating these large aggregate systems quantum mechanically, the full significance of intersite correlation has not been yet known.
 
 In this work, we apply correlation through convolution.\cite{Abramavicius2011} Any correlation functions that  strictly decreases with distance can be studied with this method. A strictly decreasing correlation function implies that its Fourier transform is positive, and the existence of the square root of the covariance matrix. In either case, we assume that the disorder covariance matrix is block circulant (as is the Hamiltonian)
\begin{equation}
     C_{ij}  = \langle \varepsilon_i \varepsilon_j \rangle / \langle \varepsilon_i^2 \rangle = e^{-r_{ij}/R}
\end{equation}
  so that it is diagonalized by a 2D plane-wave Fourier-transform matrix, with eigenvalues denoted by $g$.
\begin{equation}
     C = \mathcal{F}^{-1} g \mathcal{F}.  
 \end{equation}
Correlated noise is then generated with convolution with $\sqrt{C}$.
 \begin{equation}
     \bm{\varepsilon} = \bm{\varepsilon_0} * \sqrt{C} = \mathcal{F}^{-1} [ \sqrt{g}\cdot \mathcal{F}[\bm{\varepsilon_0}]]
 \end{equation}
 and $\bm{\varepsilon_0}$ is the initial uncorrelated normal disorder with standard deviation $\sigma$.
 
 In the infinite space limit, $\sqrt{g}$ is the square root of the Hankel transform of the exponential decay $\approx \sqrt{\frac{2\pi}{Rlw}} (R^{-2} + |\bm{k}|^2)^{-3/4}$.  For small correlation lengths it is better to numerically FFT the desired convolution matrix, rather than simply use the infinite lattice functional form of $\sqrt{g}$, to avoid edge effects in the correlation.

\section{Results}
   Through a series of simple applications we show the power of a stochastic approach in describing molecular aggregates. Our studies include a scan of the point dipole coupling function parameter space in Fig. \ref{slipfig}, efficiently reproducing the earlier deterministic results of Chuang et al.\cite{Chuang2019}

   Fig. \ref{time} demonstrates the speed of the method for very large systems. The stochastic method has a roughly constant cost for small systems (where the time is dominated by the cost of extracting the Chebyshev coefficients), and the cost only rises mildly once $N$ is beyond a thousand side. While Fig. \ref{time} shows the same calculation for a fixed number of stochastic samples, the true scaling is better than linear due to self-averaging, i.e., fewer stochastic orbitals are needed for larger systems to achieve the same level of stochastic error in $\rho(\omega)$ and $\mathcal{P}(\omega)$.

\begin{figure}
    \centering
    \includegraphics[width=0.48\textwidth]{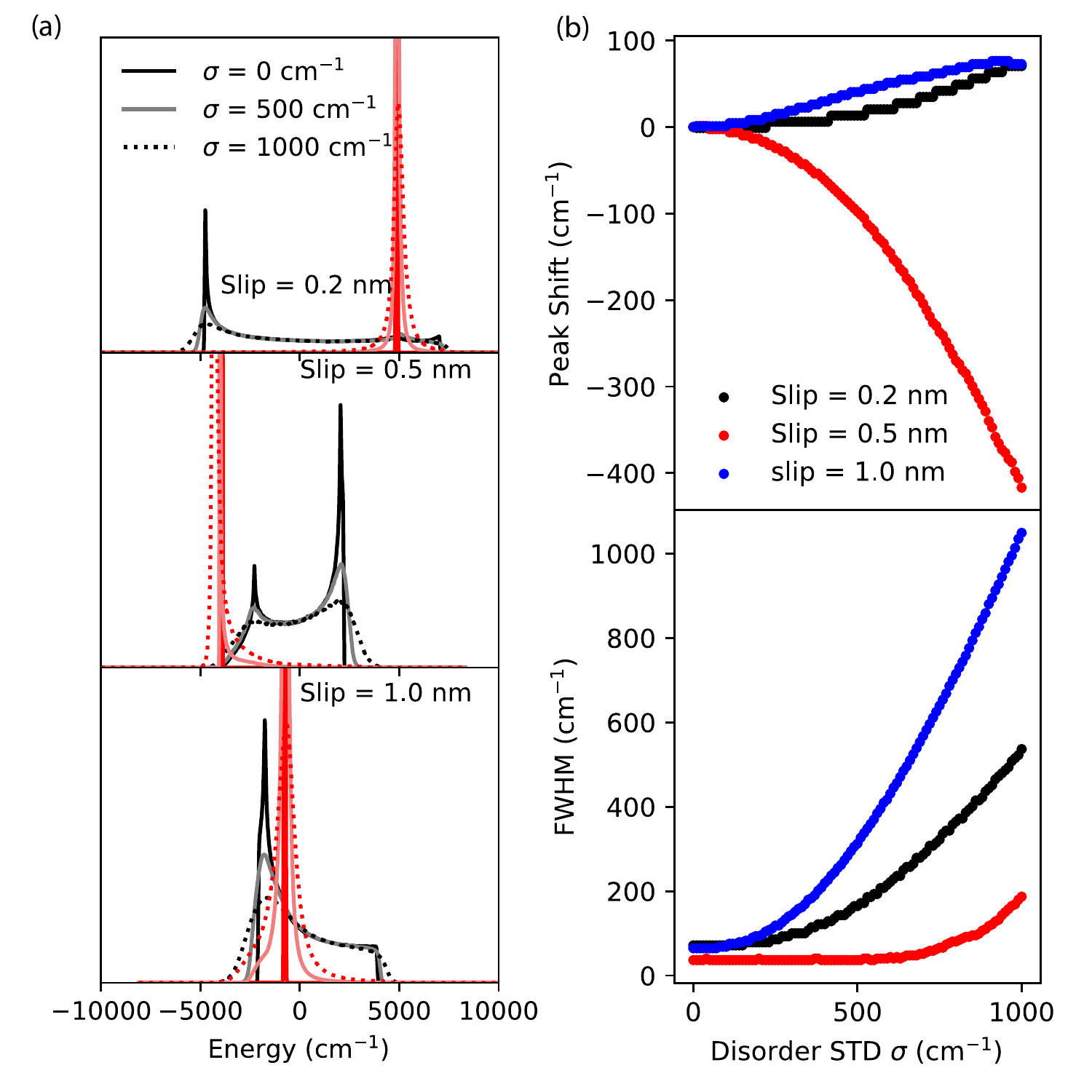}
    \caption{(a) Example density of states (colors) and absorption spectra (reds) for a H, J and I planar aggregates, with the same system setup as in Fig.\ref{slipfig}. (b) The maximum absorption peak shift and FWHM of a  H, J, and I planar aggregates of Slip = 0.2, 0.5 and 1.0 nm respectively. For the Slip = 0.5 nm band-edge J aggregate, a scaling power law of $\mathrm{FWHM} \propto \sigma^{2.1}$ was observed. }
    \label{moments}
\end{figure}

Simulating a single ``sampling'' of a typical 2D aggregate with half a million monomers, as in Fig. \ref{slipfig}, takes a mere five wall minutes on a single node. Ten such stochastic samplings (each on its own core) are sufficient for converging the DOS and absorption cross section with the full effect of disorder to within a percent at each frequency. Each of these samplings uses a different stochastic vector $\zeta$ and a different diagonal energies. Such a system is about two order of magnitudes larger than systems that could be studied with numerical diagonalization on any current computing system. Whether it be geometry, or disorder (Fig. \ref{moments}), a key point of the demonstrated application of this method is the ease of screening through parameter space. 

  In Fig. \ref{moments} we track the width and position of the absorption spectra at varying magnitudes of on-site disorder (without correlation). Our method produces non-linearities in the peak width that are similar to previous 2D tubular simulations\cite{Doria2018,Bondarenko2020} and well established scalings for 1D Kasha aggregates.\cite{Malyshev2001} Since the power law exponent scaling of the width is sensitive to the underlying geometry (slip), this method may be used as a tool for designing aggregates for particular optical properties.\cite{Thimsen2017,Bricks2015}.
  \begin{figure}
    \centering
    \includegraphics[width=0.48\textwidth]{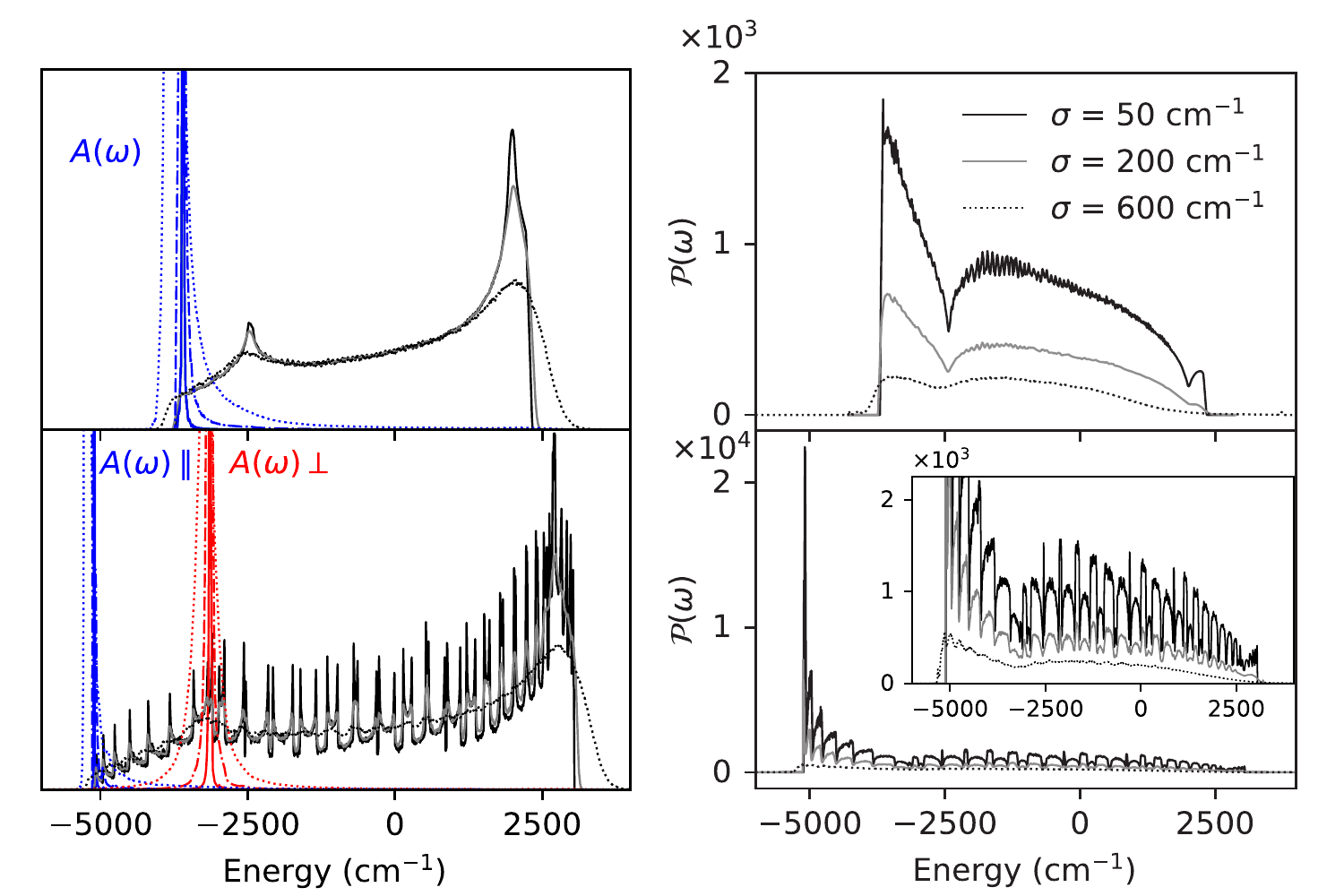}
    \caption{Density of states (left) and participation  ratios (right) for macroscopic systems at three different levels of disorder for a band edge planar aggregate (top) and the equivalent tubular aggregate (bottom). $N \approx 5\times 10^{4}$. For the tubular aggregate, a low disorder value, 50 $\mathrm{cm^{-1}}$, is not strong enough to destroy a fully delocalized bright state, while the planar aggregate is not able to support such a delocalized state. These calculations were performed with $\gamma = 2 \, \mathrm{cm^{-1}}$, and have not been interpolated to the $\gamma \to 0$ limit.}
    \label{ipr_examples}
\end{figure} 

Moving beyond the kernel approach for absorption spectra and density of states, we show in Fig. \ref{ipr_examples}  the participation ratio for large aggregates with both tubular and planar geometry.  This is the first simulation that can access an eigenvector-based observable like the participation ratio for very large systems, and also the largest participation ratio calculations for molecular aggregate systems.  The figure shows that the tubular geometry is able to support a largely delocalized bright state at the higher levels of disorder of 50-200 $\mathrm{cm^{-1}}$, while such a state is not observed in a planar aggregate for those parameters. Controlling the system localization is important for potential applications of these aggregates as photo-emitters,\cite{Hansen2008} and this work is merely a beginning for full exploration of the model space with the stochastic approach. 

\begin{figure*}
    \centering
    \includegraphics[width=0.75\textwidth]{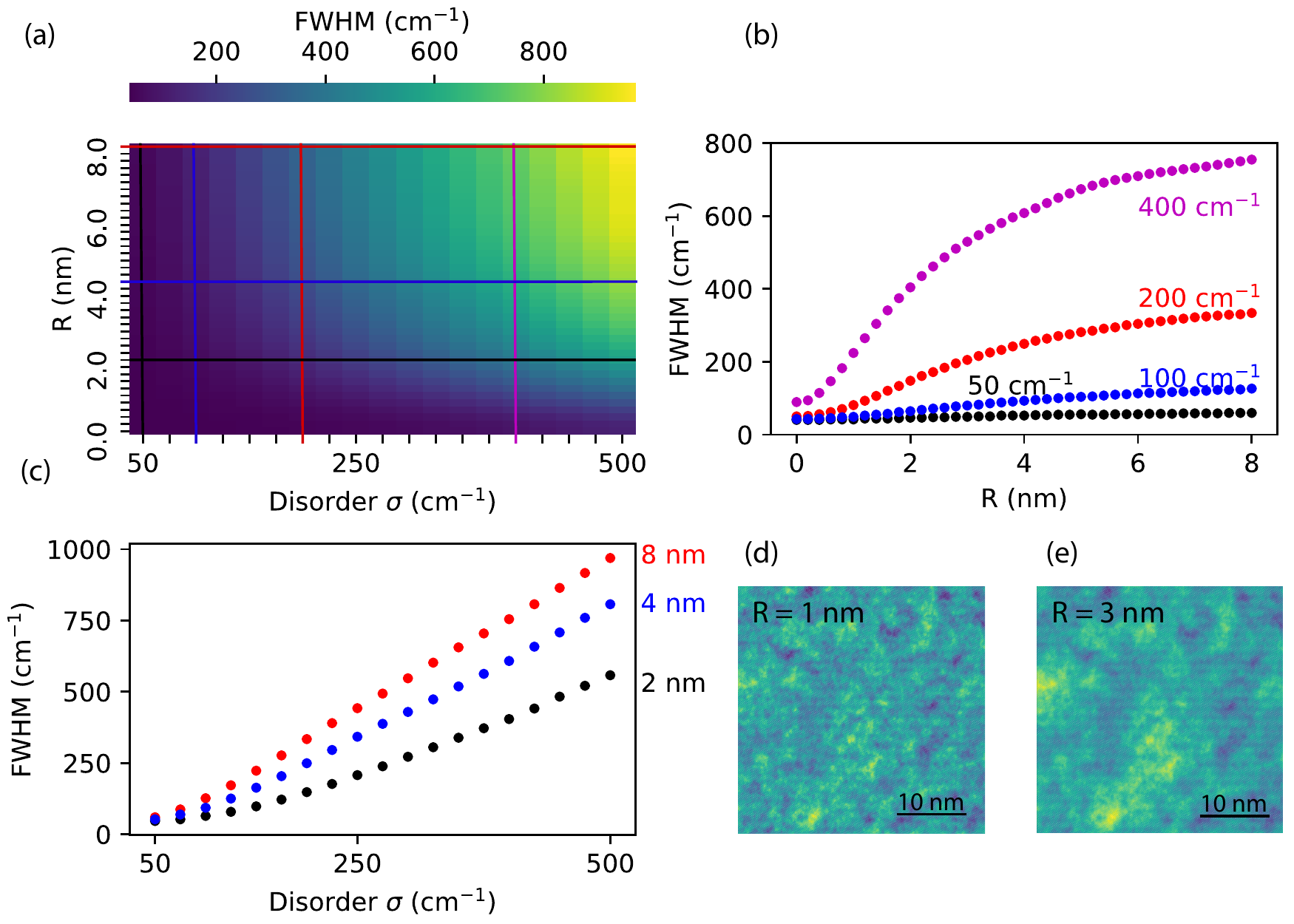}
    \caption{The width (FWHM) of the absorption spectra at varying disorder strengths and exponential correlation lengths. The full parameter space is mapped out in (a), while curves of constant disorder are shown in (b), and constant correlation (c).  (d) and (e) show an instance of exponentially correlated disorder at two different correlation lengths, as generated by the same random seed. We observe that $\sigma$ is not a separable variable from $R$, and a more complicated re-normalization is occurring. Calculations were done on a square planar aggregate of $N = 243*1215 \approx 2 \times 10^5$ corresponding to a real space side length of $48.6 nm$.}
    \label{RcorFWHM}
\end{figure*}

In Fig. \ref{RcorFWHM}, we apply correlated disorder to a 2D planar aggregate and track the properties of the absorption spectra, fully mapping out the disorder strength and correlation space. This figure demonstrates that even small correlation lengths extending over just a few monomers can have a drastic effect of the observed width on the absorption spectrum. Previous studies on the effect of local inter-site correlation in 1D molecular aggregates has discussed the change to absorption width in terms of the small-N phenomena of motional narrowing.\cite{Knoester1993,Knapp1984} Given how different the landscape and coupling of the 2D aggregate systems is compared with 1D and the change to the large N limit,\cite{Deshmukh2019} a new mechanism is needed to explain the effect of short length correlation.

 \section{Conclusion}
   This work shows that a stochastic approach rapidly yields the DOS, absorption, and participation ratio for large and disordered molecular aggregate systems over the full range of frequencies. We demonstrated the ability to efficiently screen the large modeling parameter space for these systems, and accurately model realistic micron-scale systems of up to a million monomers with the ability to extend to even larger systems if needed. A new stochastic approach was introduced to model delocalization via the participation ratio, going beyond previous work with the DOS.
   
   This work adds to the current knowledge of 2D and tubular molecular aggregates. We map out the entirety of the parameter space due to varying the lattice angle (Slip), and the effects of disorder and correlated disorder on the optical spectrum. We find that the effect of correlation on the absorption is strong even at short length scales, and is not separable from the strength of the disorder.
   
   Future extensions of the stochastic method presented here would tackle more challenging dynamic optical properties that are not be feasible for large systems with a deterministic approach. Sample applications include  time-dependent treatment of exciton lifetime, coherences, and diffusion,\cite{Chuang2016}, system environment and vibronic bath effects,\cite{Pandya2018}, or a multi-excitonic basis looking at transport and recombination properties.\cite{Tempelaar2017} 

\section*{Acknowledgements}
DN is grateful for support by NSF grant CHE-1763176. Computational resources were supplied through the XSEDE allocation TG-CHE170058.  In addition, DN and RB gratefully acknowledges the support from the US-Israel Binational Science Foundation (BSF) under Grant No. 2018368. JRC thanks the support of NSF CHE-190524 grant. APD thanks UCLA Chemistry and Biochemistry Excellence in Research Fellowship and SG Fellowship for financial support. ER acknowledges support from the Department of Energy, Photonics at Thermodynamic Limits Energy Frontier Research Center, under Grant No. DE-SC0019140. This paper was supported by the Center for Computational Study of Excited State Phenomena in Energy Materials (C2SEPEM), which is funded by the U.S. Department of Energy, Office of Science, Basic Energy Sciences, Materials Sciences and Engineering Division via Contract No. DE-AC02-05CH11231, as part of the Computational Materials Sciences Program. 

\appendix
\section{Geometric Parameters \label{geometry}}

    For all figures, unless otherwise specified, we use a planar aggregate with brick size $(l,w) = (2.0 \, \mathrm{nm}, 0.4 \, \mathrm{nm} )$ with dipoles pointing out of the plane at a zenith angle of $\phi = 70^{\circ}$, $\bm{\mu}_n = (0, \sin(\phi), \cos(\phi)) $,  as done by Chuang et al.\cite{Chuang2019} Furthermore, if no slip was given, the standard structure will be a band edge J-aggregate
    of $\textrm{Slip} =0.5$ nm. As such, a lattice point is generated by 
    \begin{equation}
         \bm{r}_n = \begin{pmatrix} x \\ y \\
        \end{pmatrix} = \begin{pmatrix} w & 0 \\ s & l \\
        \end{pmatrix} \begin{pmatrix} i_x \\ i_y \\
        \end{pmatrix}
    \end{equation}
    and $i_x = 1\cdots N_x$, $i_y = 1 \cdots N_y$. For the purpose of disorder, the Fourier modes are then generated by 
    \begin{equation}
        \bm{k_n} = \begin{pmatrix}
            k_x \\ k_y\\
        \end{pmatrix} = 
        \begin{pmatrix}
        \frac{2\pi}{wN_x}  & \frac{- \pi s}{lwN_y}\\
        0 & \frac{2 \pi}{l N_y} \\
        \end{pmatrix}
        \begin{pmatrix}
            \tilde{i}_{kx}\\ \tilde{i}_{ky} \\
        \end{pmatrix}
    \end{equation}  
    where the indices of $\tilde{i}_{k}$ are correctly wrapped around the periodic boundaries,  such that $\bm{k}$ is in the first Brillouin zone and has the smallest possible norm.
    
    For the tubular aggregate figures, a band edge J-aggregate parameters are used as given by Didraga et al.\cite{Didraga2002,Didraga2004} The parameters used generate an equivalent tubular aggregate to a planar aggregate with 0.5 nm slip and a (3,3) chiral vector. A radius of 5.4553 nm is used, with the dipole angle relative to the plane being $\beta = 47.4^{\circ}$, a height between rings of 0.0467 nm, between ring rotation of $\delta= 6.7^{\circ}$ and with rotational symmetry of $N_r = 2$. The common ``herringbone" structural model for tubular aggregates was not studied in this paper. 
\begin{figure}
    \centering
    \includegraphics[width=0.48\textwidth]{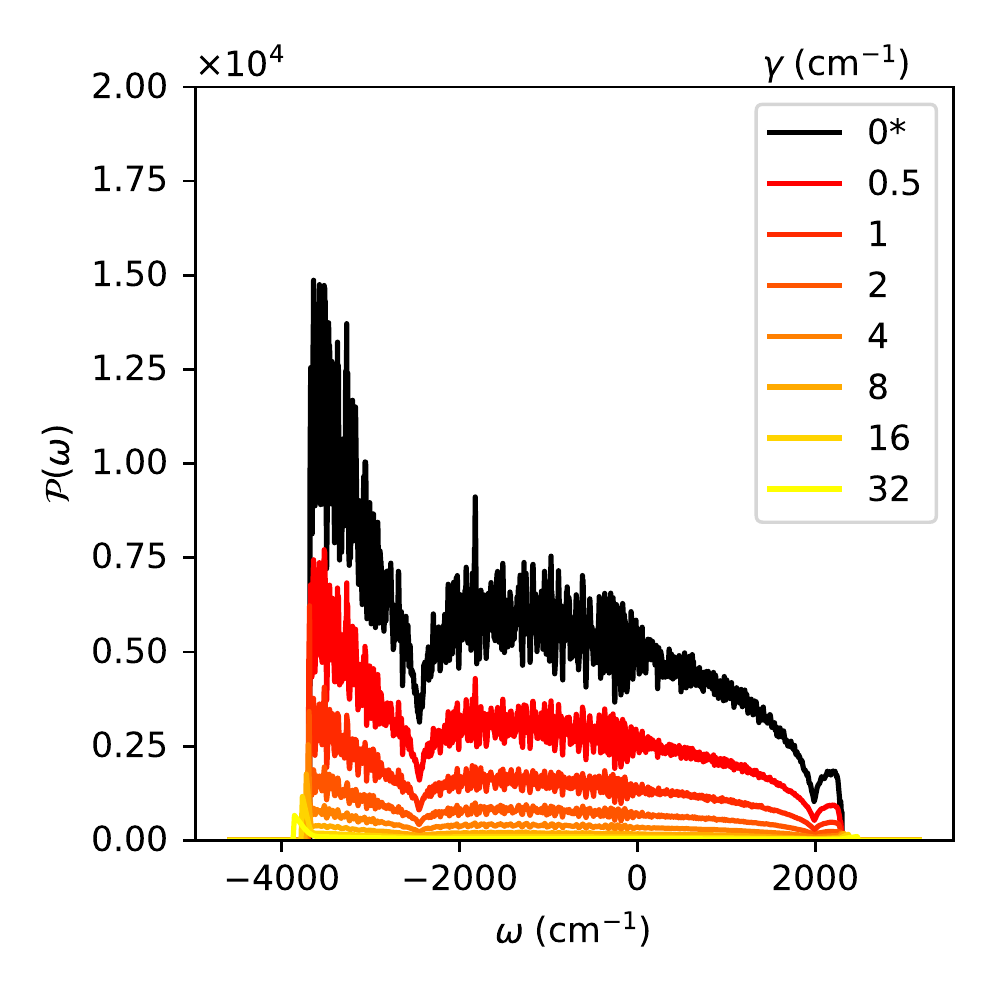}
    \caption{Stochastic participation ratio at varying line broadening parameters, $\gamma$. Vandermonde polynomial interpolation is used at each data point to reach the $\gamma \xrightarrow[]{} 0 $ limit, and is shown by the black curve. This calculation was done for a planar 2D aggregate of $N \approx 5\times 10^4$, as in Fig. \ref{ipr_examples}, with 50 $\mathrm{cm^{-1}}$ of static disorder. All spectra are generated by the same random seed. }
    \label{interpolation}
\end{figure}
\section{Participation Ratio, and Interpolation to Delta Functions \label{interpolated}}

Even for fairly large systems, low amounts of static disorder can cause a constant error in the stochastic participation ratio due to the Gaussian regularization of the delta function kernel, and numerical degeneracy in the eigenstates. For example, when there is a high amounts of static disorder, such as in Fig. \ref{ipr.small} where $\sigma/(\max (J) )$ is approximately half, we see convergence to the true matrix-diagonalized participation ratio even at fairly large $\gamma$. However for smaller amounts of disorder, such as when $\sigma/(\max (J) )$ is less than 10 percent, degeneracy in the eigenvalues becomes an issue for stochastic sampling. 

To address this,  in Fig. \ref{interpolation} we sample a large system at varying degrees of line-broadening and find that $\mathcal{P} ( \omega) \propto \frac{1}{\gamma}$, as would naturally be suggested by the functional form of the Gaussian limit of the delta function. This form of error was found to be constant across the band. Using Vandermonde polynomial interpolation, we reconstruct an approximation for the true delta function limit at $\gamma \xrightarrow[]{} 0$. There is a  constant error across the band introduced by insufficiently small gamma, which is independent of the geometry of the system.  Convergence of the interpolation suggests  that the stochastic participation ratio converges to a finite value that is lower than its theoretical bound of $N$. The constant error  suggests that the line-shapes for the stochastic participation ratio are correct, so comparisons between different systems at the same line broadening are valid.

\section{Stochastic Absorption beyond the Dipole Approximation \label{fullabsorb}}

Calculating the absorption beyond the dipole approximation requires filtering of the collective dipoles of each exciton to obtain the eigenstate at a particular wavevector $\textbf{k}$. Stochastically, we will extract the $k$-dependent information by starting with spatially random state and filtering them, spatially, after the frequency filtering, i.e.,
\begin{equation}
    A_{\bm{k}}(\omega) \propto \Big\{
    \bra{\zeta} \, \bm{\mu\cdot\epsilon} \,
    P_{\bm{k}} \,
    \delta(H - \omega)\,
    \bm{\mu\cdot\epsilon}
    \ket{\zeta}\Big\},
\end{equation}
where $P_{\bm{k}}=
\ket{\bm{k}}\bra{\bm{k}}$
is a spatial filter at the wavevector
$\bm{k}$.
 Thus, we will apply a delta Chebyshev filter to select for frequency-selected eigenstates followed by a spatial filter that selects for overlap with the applied wavevector of the radiation. Dichroism can similarly be extracted as we do under the dipole approximation in the main section.

\bibliography{main}

\end{document}